\documentclass[preprint,12pt]{elsarticle}

\usepackage{graphicx}

\usepackage{amssymb}

\journal{Nuclear Physics A}

\begin{document}

\begin{frontmatter}



\title{Investigation of double beta decay of $^{100}$Mo to excited states of $^{100}$Ru}




\author[IPHC]{R.~Arnold}
\author[LAL]{C.~Augier}
\author[ITEP]{A.S.~Barabash\corref{cor1}}
\author[UCL]{A.~Basharina-Freshville}
\author[LAL]{S.~Blondel}
\author[UM]{S.~Blot}
\author[LAL]{M.~Bongrand}
\author[JINR]{V.~Brudanin}
\author[CPPM]{J.~Busto}
\author[INL]{A.J.~Caffrey}
\author[IEAP]{P.~\v{C}erm\'{a}k}
\author[UB]{C.~Cerna}
\author[LPC]{A.~Chapon}
\author[UM]{E.~Chauveau}
\author[NRPI]{L.~Dragounov\'{a}}
\author[LAPP]{D.~Duchesneau}
\author[LPC]{D.~Durand}
\author[JINR]{V.~Egorov}
\author[LAL,UCL]{G.~Eurin}
\author[UM]{J.J.~Evans}
\author[UCL]{R.~Flack}
\author[LAL]{X.~Garrido}
\author[LAL]{H.~G\'omez}
\author[LPC]{B.~Guillon}
\author[UM]{P.~Guzowski}
\author[IEAP]{R.~Hod\'{a}k}
\author[UB]{P.~Hubert}
\author[UB]{C.~Hugon}
\author[NRPI]{J.~H\accent23ulka}
\author[LAL]{S.~Jullian}
\author[JINR]{A.~Klimenko}
\author[JINR]{O.~Kochetov}
\author[ITEP]{S.I.~Konovalov}
\author[JINR]{V.~Kovalenko}
\author[LAL]{D.~Lalanne}
\author[UTA]{K.~Lang}
\author[LPC]{Y.~Lemi\`{e}re}
\author[UTA]{Z.~Liptak}
\author[LSM]{P.~Loaiza}
\author[UB]{G.~Lutter}
\author[IEAP]{F.~Mamedov}
\author[UB]{C.~Marquet}
\author[LPC]{F.~Mauger}
\author[UW]{B.~Morgan}
\author[UCL]{J.~Mott}
\author[JINR]{I.~Nemchenok}
\author[OU]{M.~Nomachi}
\author[UTA]{F.~Nova}
\author[IPHC]{F.~Nowacki}
\author[Saga]{H.~Ohsumi}
\author[UTA]{R.B.~Pahlka}
\author[UB]{F.~Perrot}
\author[UB,LSM]{F.~Piquemal}
\author[FMFI]{P.~Povinec}
\author[UW]{Y.A.~Ramachers}
\author[LAPP]{A.~Remoto}
\author[LSCE]{J.L.~Reyss}
\author[UCL]{B.~Richards}
\author[INL]{C.L.~Riddle}
\author[IEAP]{E.~Rukhadze}
\author[JINR]{N.~Rukhadze}
\author[UCL]{R.~Saakyan}
\author[LAL]{X.~Sarazin}
\author[IMP,JINR]{Yu.~Shitov}
\author[LAL,IUF]{L.~Simard}
\author[FMFI]{F.~\v{S}imkovic}
\author[IEAP]{A.~Smetana}
\author[IEAP]{K.~Smolek}
\author[JINR]{A.~Smolnikov}
\author[UM]{S.~S\"{o}ldner-Rembold}
\author[UB]{B.~Soul\'{e}}
\author[IEAP]{I.~\v{S}tekl}
\author[Jyv]{J.~Suhonen}
\author[MHC]{C.S.~Sutton}
\author[LAL]{G.~Szklarz}
\author[UCL]{J.~Thomas}
\author[JINR]{V.~Timkin}
\author[UCL]{S.~Torre}
\author[JINR]{V.I.~Tretyak}
\author[INR]{Vl.I.~Tretyak}
\author[ITEP]{V.~Umatov}
\author[UCL]{C.~Vilela}
\author[CUP]{V.~Vorobel}
\author[LSM]{G.~Warot}
\author[UCL]{D. Waters}
\author[CUP]{A.~\v{Z}ukauskas}

\cortext[cor1]{Corresponding author}

\address{\vspace{0.1 in}(The NEMO--3 Collaboration)\vspace{0.05 in}}

\address[IPHC]{IPHC, UPL, CNRS/IN2P3, F-67037 Strasbourg, France}
\address[LAL]{LAL, Univ Paris-Sud, CNRS/IN2P3, F-91405 Orsay, France}
\address[ITEP]{ITEP, Institute of Theoretical and Experimental Physics, 117218 Moscow, Russia}
\address[UCL]{University College London, London WC1E 6BT, United Kingdom}
\address[UM]{University of Manchester, Manchester M13 9PL, United Kingdom}
\address[JINR]{JINR, Joint Institute for Nuclear Research, 141980 Dubna, Russia}
\address[CPPM]{CPPM, Universit\'{e} de Marseille, CNRS/IN2P3, F-13288 Marseille, France}
\address[INL]{Idaho National Laboratory, Idaho Falls, ID 83415, U.S.A.}
\address[IEAP]{IEAP, Czech Technical University in Prague,  CZ-12800 Prague, Czech Republic}
\address[UB]{CENBG, Universit\'{e} Bordeaux, CNRS/IN2P3, F-33175 Gradignan, France}   
\address[LPC]{LPC Caen, ENSICAEN, Universit\'{e} de Caen, CNRS/IN2P3, F-14050 Caen, France}
\address[NRPI]{National Radiation Protection Institute, CZ-14000 Prague, Czech Republic}
\address[LAPP]{LAPP, Universit\'{e} de Savoie, CNRS/IN2P3, F-74941 Annecy-le-Vieux, France}
\address[UTA]{University of Texas at Austin, Austin, TX 78712, U.S.A.}
\address[LSM]{Laboratoire Souterrain de Modane CNRS/CEA, F-73500 Modane, France}
\address[UW]{University of Warwick, Coventry CV4 7AL, United Kingdom}
\address[OU]{Osaka University, 1-1 Machikaney arna Toyonaka, Osaka 560-0043, Japan}
\address[Saga]{Saga University, Saga 840-8502, Japan}
\address[FMFI]{FMFI, Comenius University, SK-842 48 Bratislava, Slovakia}
\address[LSCE]{LSCE, CNRS, F-91190 Gif-sur-Yvette, France}
\address[IMP]{Imperial College London, London SW7 2AZ, United Kingdom}
\address[IUF]{Institut Universitaire de France, F-75005 Paris, France}
\address[Jyv]{Jyv\"{a}skyl\"{a} University,  FIN-40351 Jyv\"{a}skyl\"{a}, Finland}
\address[MHC]{MHC, South Hadley, Massachusetts 01075, U.S.A.}
\address[INR]{Institute for Nuclear Research, MSP 03680, Kyiv, Ukraine}
\address[CUP]{Charles University in Prague, Faculty of Mathematics and Physics, 
CZ-12116 Prague, Czech Republic}

\begin{abstract}

Double beta decay of $^{100}$Mo to the excited states 
of daughter nuclei
has been studied using a 600 cm$^3$ low-background HPGe detector and 
an external 
source consisting of 2588 g of 97.5\% enriched metallic $^{100}$Mo, which was
formerly inside the NEMO-3 detector and used for the NEMO-3 measurements of $^{100}$Mo. The half-life for 
the two-neutrino 
double beta decay of $^{100}$Mo to the excited 0$^+_1$ state in $^{100}$Ru is 
measured to be 
$T_{1/2}=[7.5 \pm{0.6}(stat) \pm {0.6}(syst)]\cdot 10^{20}$ yr. 
For other $(0\nu + 2\nu)$ transitions to the
2$^+_1$, 2$^+_2$, 0$^+_2$, 2$^+_3$ and 0$^+_3$ levels in $^{100}$Ru,  limits 
are obtained at the level of 
$\sim (0.25-1.1)\cdot 10^{22}$ yr. 

\end{abstract}


\end{frontmatter}


\section{Introduction}
\label{}

Experiments with solar, atmospheric, reactor and accelerator neutrinos 
have provided compelling 
evidence for the existence of neutrino oscillations driven by non zero 
neutrino masses and neutrino mixing \cite{CLE98,FUK96,ANS92,ABD09,AHM01,FUK02,
FUK98,ASH04,EGU03,ARP08,AHN06,MIC06} (see also reviews 
\cite{MOH06,BIL11,VAL12}).
The detection and study of neutrinoless double beta 
($0\nu\beta\beta$)
decay may clarify the following problems of neutrino physics (see discussions 
in \cite{MOH05,PAS06,VER12,BIL12,ROD12,SIM13}): (i) neutrino nature: 
whether the neutrino is a Dirac or a Majorana particle, (ii) absolute neutrino 
mass scale, (iii) the type of neutrino mass
hierarchy (normal, inverted, or quasidegenerate), (iv) CP violation in the 
lepton sector (measurement of the Majorana CP-violating phases).

Double
beta decay with the emission of two neutrinos $(2\nu\beta\beta)$ is
an allowed process of second order in the Standard Model. 
The $2\nu\beta\beta$ decays provide the
possibility of an experimental determination  of the 
nuclear matrix  elements (NME) involved
in the double beta decay  processes.  This leads to the development of
theoretical  schemes for NME calculations  both in
connection   with  the   $2\nu\beta\beta$  decays   as  well   as  the
$0\nu\beta\beta$ decays (see, for example, \cite{ROD06,KOR07,KOR07a,SIM08,IAC13}).  
At present,
$2\nu\beta\beta$ decay  to  the ground  state  of the  final
daughter   nucleus  has been   measured  for eleven  nuclei:  $^{48}$Ca,
$^{76}$Ge, $^{82}$Se, $^{96}$Zr, $^{100}$Mo, $^{116}$Cd, $^{128}$Te, 
$^{130}$Te, $^{136}$Xe, 
$^{150}$Nd  and $^{238}$U  (a review of the results is given in
Ref.~\cite{BAR10}, for $^{136}$Xe see recent results of EXO \cite{AUG12} and KamLAND-Zen \cite{GAN12}).
In addition two neutrino double electron capture was detected in $^{130}$Ba \cite{MES01,MAG09}.

The $\beta\beta$  decay can proceed through transitions  to the ground
state as  well as to various  excited states of  the daughter nucleus.
Studies of  the latter transitions  allow one to obtain supplementary
information about  $\beta\beta$ decay.  
Because of the smaller transition
energies,  the  probabilities  for $\beta\beta$ decay  to
excited  states  are   substantially  suppressed  in  comparison  with
transitions to the ground state,
but as it was shown in Ref.~\cite{BAR90}, by
using   low-background High Purity Germanium
(HPGe)   detectors,  the 
$2\nu\beta\beta$ decay  to the $0^+_1$  level in the  daughter nucleus
may  be detected for  such nuclei  as $^{100}$Mo,  $^{96}$Zr, and
$^{150}$Nd.  For these isotopes  the energies involved in the
$\beta\beta$ transitions are large enough (1904,  2202, and
2631~keV, 
respectively),  and  the  expected  half-lives  are of  the  order  of
$10^{20} - 10^{21}$~yr. 
The double beta decay of $^{100}$Mo to 
the 0$^+$ excited state at 1130.3 keV of $^{100}$Ru was first observed in 
\cite{BAR95}, and later confirmed in independent experiments 
\cite{BAR99,DEB01,HOR06,ARN07,KID09,BEL10}. In 2004,
the transition was detected in $^{150}$Nd for the first time
\cite{BAR04,BAR09a}. 
For $^{96}$Zr only a limit has been obtained up to now ($T_{1/2} > 6.8\cdot 10^{19}$ yr \cite{BAR96}).
Additional    isotopes ($^{82}$Se,    $^{130}$Te,
$^{116}$Cd, and $^{76}$Ge) have also become of interest to
studies of the $2\nu\beta\beta$ decay to the $0^+_1$ level (see reviews
in Refs.~\cite{BAR00,BAR07,BAR10a}).

Recently it was speculated \cite{DOL05} that neutrinos may violate the Pauli exclusion principle (PEP)
and therefore, at least partly obey Bose-Einstein statistics (see also \cite{CUC96}). As a consequence,
neutrinos could form a Bose condensate which may account for parts or even all of the dark
matter in the universe. As discussed in \cite{BAR07a} the possible violation of the PEP has interesting
consequences for $2\nu\beta\beta$ decay. It not only modifies the energy and angular distributions of the
emitted electrons, but it also strongly affects the $2\nu\beta\beta$ decay rates to the ground 
and excited states in
daughter nuclei. Following \cite{BAR07a}, the half-life ratios for transitions to excited $2^+$ states and
the $0^+$ ground state are by far the most sensitive way of obtaining bounds on a
substantial bosonic component to neutrino statistics. As a result, information on the decay
rates to excited states is needed to test this new and potentially far-reaching hypothesis.

The $0\nu\beta\beta$ transition  to excited
states of  daughter nuclei provides a unique signature:
in addition to two electrons with fixed
total energy, one  ($0^+ \to 2^+_1$ transition) or two ($0^+
\to  0^+_1$ transition) photons appear,  with their energies  being strictly  fixed. 
In  a hypothetical experiment  detecting all decay products  with a
high efficiency and a high  energy resolution, the background can be
reduced to nearly zero.  This zero background idea will be the goal of
future experiments featuring a large mass of the $\beta\beta$ sample
(as mentioned in Refs.~\cite{BAR00,BAR04a,SUH00}).     
In Ref.~\cite{SIM02} it  was mentioned that detection  of this transition
will also give us  the additional   possibility of distinguishing between the various
$0\nu\beta\beta$ mechanisms (the light and heavy Majorana neutrino
exchange mechanisms, the trilinear R-parity breaking mechanisms etc.).

\begin{figure*}
\begin{center}
\includegraphics[width=12.8cm]{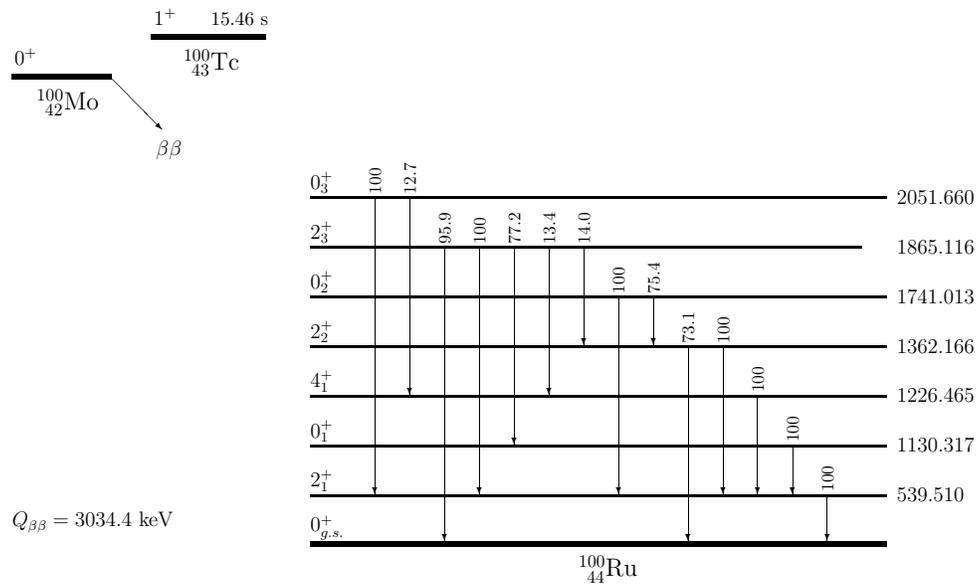}
\caption{\label{fig:fig1}The decay scheme of $^{100}$Mo, taken 
from \cite{SIN08}. 
Only the investigated levels of $0^+$ and $2^+$ and  4$^+_1$ level associated 
with investigated transitions
are shown. The relative main branching ratios from each level are 
presented.}  
\end{center}
\end{figure*}

In this article, results of an experimental investigation
of the $\beta\beta$ decay of $^{100}$Mo to the excited states of
$^{100}$Ru are presented. 
The decay scheme for the triplet $^{100}$Mo - $^{100}$Tc - $^{100}$Ru  
\cite{SIN08} is shown in Fig.~\ref{fig:fig1}. 
The measurements have been carried out using a
HPGe detector to look for $\gamma$-ray lines corresponding to 
the decay scheme. We study here transitions up to the 2051.7 keV level
of $^{100}$Ru. Smaller phase space factors and thus smaller probabilities
make it less useful to study the higher excited levels.

\section{Experimental study}

The experimental work was performed in the Modane Underground
Laboratory (depth of 4800 m w.e.). A 600 cm$^3$ low-background HPGe detector 
was used to measure a 2588 g sample of enriched $^{100}$Mo metallic foil 
in a special delrin box which was placed around the detector endcap.
This sample was formerly inside the NEMO-3 detector and used for the NEMO-3 
measurements of $^{100}$Mo \cite{ARN05,ARN07}. 
Taking into account the concentration of Mo (99.8\%) and enrichment (97.5\%) there were  
2518 g of $^{100}$Mo (or $1.52\cdot10^{25}$ nuclei of $^{100}$Mo) in the sample. 
Data were collected for 2288 h.

The Ge spectrometer is composed of a p-type crystal.  
The cryostat, endcap, and the other mechanical parts are made of 
a very pure Al-Si alloy. The cryostat has a U-type geometry to shield the crystal from 
radioactive impurities in the dewar. The passive shielding consists of three layers of
Roman lead (which can be removed to host different sample volumes) 
with a total thickness of $\sim 12$ cm  and an external layer of $\sim 20$ cm of low radioactivity lead. 
The activity of Roman lead is below 100 mBq/kg and the activity of low radioactivity lead is $\sim$ 
5-20 Bq/kg.
To remove $^{222}$Rn gas, a special 
effort was made to minimize the free space near the detector and, in addition, 
the passive shielding is flushed with radon-depleted air (concentration of 
$^{222}$Rn is $\sim$ 15 mBq/m$^3$) from a radon trapping facility.

The electronics consist of currently available spectrometric amplifiers and
a 16384 channel ADC. The energy calibration is adjusted to cover the energy 
region from 5 keV to 3.5 MeV (the detector is sensitive to energies up to 6 MeV), and the energy resolution is
2.0 keV for the 1332-keV line of $^{60}$Co. The electronics are stable during
the experiment due to the constant conditions in the laboratory (temperature 
of $\approx 23^\circ$ C, hygrometric degree of $\approx 50$\%).  

All necessary information about radioactive isotopes was taken  
from 
databases of the National Nuclear Data Center \cite{NNDC} 
and were used
for analysis of the energy spectrum. 
The photon detection efficiency for each investigated process 
has been calculated
with the Monte Carlo (MC) code GEANT 3.21 \cite{GEANT3} (and re-checked with GEANT 4 \cite{GEANT4}). 

To increase the accuracy of the efficiency calculations, special calibration measurements 
using radioactive sources with well-known activity ($^{238}$U, $^{152}$Eu and $^{138}$La) have 
been carried out.

The uranium source, with a diameter of 47 mm and height of 3 mm, contained 5.74 g of uranium 
ore (IAEA-RGU-1 reference material \cite{IAEA}) with an activity of (28.36 $\pm$ 0.09) Bq. Two measurements 
were carried out: in the first case the source was placed directly on the endcap (in the center) 
of the HPGe detector, and in the second, it was shifted by 33 mm above the endcap. More than 10 
different energy gamma-rays 
(from $^{214}$Pb and $^{214}$Bi) in the region (100-2500) keV were used for the
calibration. 

The europium source was a point-like source with an activity (2323 $\pm$ 46) Bq. The source was 
placed 310 mm above the endcap.
More than 10 different energy gamma-rays in the region (122-1408) keV were used for the calibration.  

In the case of the lanthanum, the source was a mixture of powdery filler (934 g of flour) 
and La$_2$O$_3$ powder (238 g). This mix was 
moulded into approximately the
same geometry and size as the box used in the $^{100}$Mo measurement.
Taking into account the abundance of $^{138}$La ((0.0888 $\pm$ 0.0007) \% \cite{SON03}) and its half-life 
($T_{1/2} = (1.02 \pm 0.01)\cdot 10^{11}$ yr 
\cite{SON03}) it is possible to calculate precisely the activity 
of $^{138}$La: (168 $\pm$ 2) Bq. Two gamma-rays are emitted with energies 788 and 1435 keV.

The results of these calibration measurements were used to check the accuracy of the MC simulations. 
The MC calculations for efficiency were adjusted to the results of the calibration  measurements 
changing some parameters of the detector (mainly by increasing the dead layer of the HPGe 
detector). As a result discrepancies between experimental and simulated efficiencies
do not exceed 7\% for all gamma lines of the 3 sources.

\begin{figure}
\begin{center}
\includegraphics[width=8.6cm]{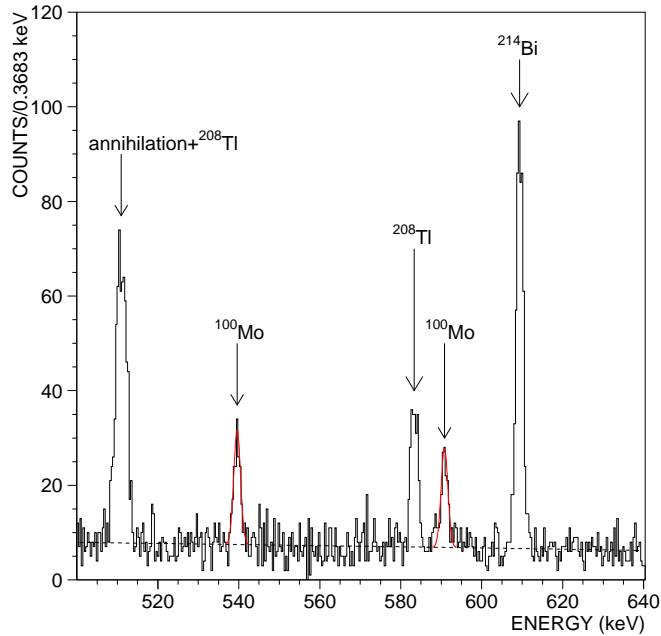}
\caption{\label{fig:fig2}Energy spectrum from enriched Mo in the range [500-640] keV. 
The dashed line is the estimated continuous background and colored lines are the 
fitted peaks at 539.5 and 590.8 keV (see text).}  
\end{center}
\end{figure}

\begin{figure}
\begin{center}
\includegraphics[width=8.6cm]{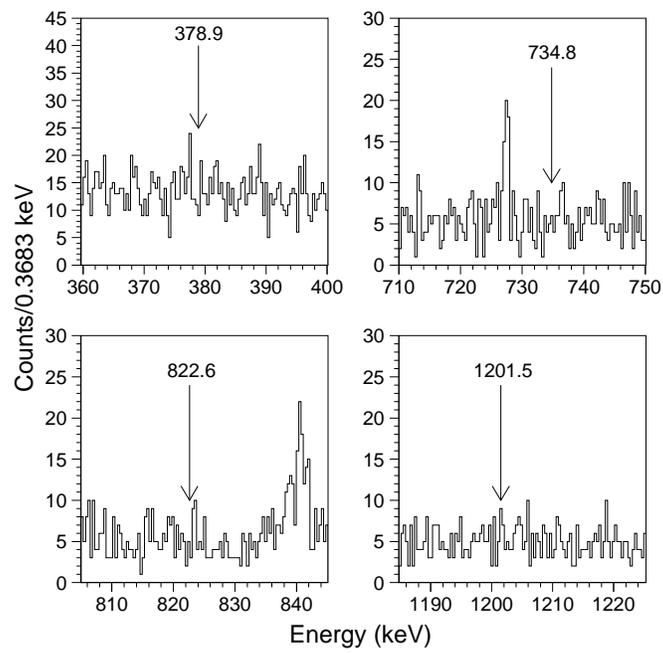}
\caption{\label{fig:fig3}Energy spectrum from enriched Mo  
in the range [360-400], [710-750], [800-850] and [1180-1230] keV.}  
\label{fig_2}
\end{center}
\end{figure}

\begin{figure}
\begin{center}
\includegraphics[width=8.6cm]{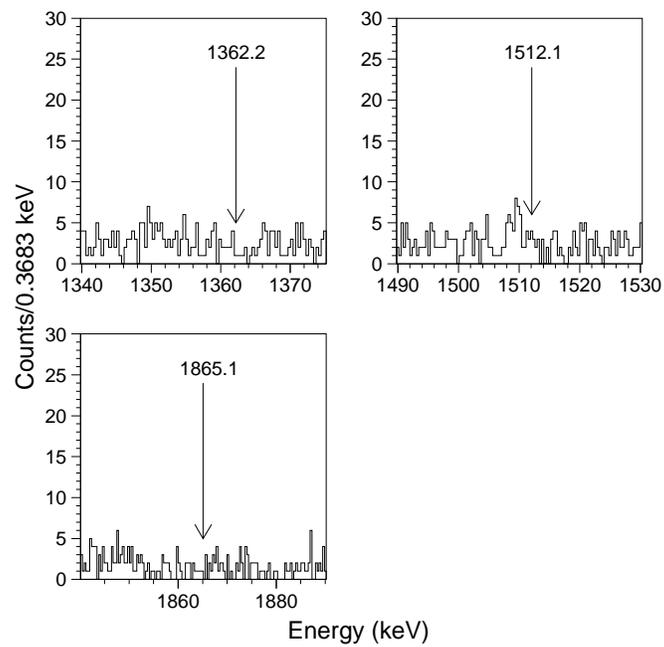}
\caption{\label{fig:fig4}Energy spectrum from enriched Mo 
in the ranges [1340-1380], [1490-1530]  and [1840-1890] keV.}  
\end{center}
\end{figure}

Figures \ref{fig:fig2}, \ref{fig:fig3}, and \ref{fig:fig4}
show the energy spectra in the ranges of interest. Arrows indicate the position of the peaks
under study. 

\section{Analysis and results}

\subsection{Decay to the 0$^+_1$ excited state}

This transition is accompanied by two $\gamma$-rays with energies of 
539.5 keV and
590.8 keV (see Fig. 1). The detection photopeak efficiencies are equal to 3.29\% 
at 539.5 keV and 3.22\% at 590.8 keV. 
These values were obtained using MC simulations.
The calculation included the effects of the extended
geometry, the attenuation of the $\gamma$-rays in the sample, the 
full-energy peak efficiency of the germanium detector, 
the summing effects in the detector and the  
anisotropic angular correlation between the $\gamma$-rays emitted after 2$\beta$ decay 
of $^{100}$Mo to the 0$^+_1$ excited state of $^{100}$Ru.
Fig.~\ref{fig:fig2} shows the energy spectrum 
in the range of interest. Both peaks at 539.5 keV and 590.8 keV are clearly visible.
In addition there are also peaks from background at 511 keV (annihilation + $^{208}$Tl), 
583.2 keV ($^{208}$Tl) and 609.3 keV ($^{214}$Bi).
The averaged continuous background (dashed line) is obtained by
fitting a parabola to the (500-640) keV energy range after removing 
counts under peaks around 511, 539.5, 583.2, 590.8 and 609.3 keV. 
Then the peaks at 539.5 and 590.8 keV are fitted by Gaussians with energy resolution equal to 2.0 keV 
(colored lines).
The average energy resolution over the entire period of measurements was 
established by fitting the neighboring 609 keV $\gamma$-line, ($2.0 \pm 0.1$) keV.
Finally,
the peak at 539.5 keV is at 129 $\pm$ 14 counts, and the peak at 590.8 keV is at
110 $\pm$ 13 counts. It corresponds to $T_{1/2}=7.0^{+0.9}_{-0.7}(stat)\cdot 10^{20}$ yr
for the 539.5 keV peak and $T_{1/2}=8.0^{+1.1}_{-0.8}(stat)\cdot 10^{20}$ yr for
590.8 keV peak.
Summing the two peaks we obtain a signal
of 239 $\pm$ 19 events, corresponding to a half-life of $^{100}$Mo 
to the first $0^+$ 
excited state of $^{100}$Ru given by 
$T_{1/2}=[7.5 \pm {0.6}(stat) \pm {0.6}(syst)]\cdot 10^{20}$ yr. 
The primary systematic comes from the efficiency calculations (7\%), 
the estimation of the number of useful events (4\%) and the uncertainties 
in the geometrical position of the $^{100}$Mo sample (2\%). The uncertainty in the 
number of useful events is connected with an error in the definition of the average background in the region of the
studied peaks and the choice of the fitting procedure. The uncertainty connected with the inaccuracy in the position of the $^{100}$Mo sample in relation to the position of the crystal was estimated 
by the MC, changing the position of the sample by $\pm$ 3 mm along a crystal axis and changing the
external diameter of the sample by $\pm$ 4 mm.

\subsection{\label{sec:level3}Decay to the 2$^+_1$ excited state}

To search for this transition, one has to look for a gamma-ray with an energy of 539.5 keV. 
For this single gamma-ray, the detection efficiency is 4.02\%. Decays to the 0$^+_1$ state also contribute 
to a peak at this energy. In the analysis given above, two measurements of the half life for decays to 
the 0$^+_1$ state are given, obtained independently from the 539.5 keV and 590.8 keV peaks. 
These two measurements are consistent with each other. Since the 590.8 keV peak contains 
only photons from decays to the 0$^+_1$ state, this shows that the observed peak at 539.8 keV 
is consistent with decays to only the 0$^+_1$ state. Therefore, one can only give a lower limit 
on the half life for transitions to the 2$^+_1$ excited state of $^{100}$Ru.
The contribution to the 539.5 keV peak from $2\nu$ decay to the 0$^+_1$ excited state was 
estimated as $112 \pm 13$ events (using the observed number of events 
in 590.8 keV peak and taking into account the difference in the efficiency).
The excess of $17 \pm 19$ counts indicates that there is no signal.
The 90\% C.L. upper limit on the number of observed events for the decay to the 
2$^+_1$ state is 43.5\footnote{This value is obtained by 
integrating the Gaussian function (with mean value 17 and sigma 19) from 0 
to 43.5 to obtain 90\%  probability.}, 
yielding a limit on the half life of $T_{1/2} > 2.5\cdot 10^{21}$ yr.
The limits obtained from other measurements, together with available data on $\beta\beta$ decay of $^{100}$Mo
are presented in Table 1.

\subsection{Decays to the 2$^+_2$, 2$^+_3$, 0$^+_2$ and 0$^+_3$ excited states}

To search for these transitions one has to look for $\gamma$-rays with 
energies of 
378.9, 734.8, 822.6, 1201.5, 1362.2, 1512.1 and 1865.1 keV (Fig.~\ref{fig:fig1}).  
As one can see from figures 
\ref{fig:fig3} and \ref{fig:fig4},
there are no statistically significant peaks at
these energies.

The Bayesian approach \cite{PDG04} has been used to estimate limits. To construct the 
likelihood function, every bin of the spectrum is assumed to have a Poisson 
distribution with its mean $\mu_i$ and the number of events equal to the 
content of the $i$th bin. The mean can be written in the general form,

\begin{equation}
\mu_i = N\sum_{m} {\varepsilon_m a_{mi}} + \sum_{k}
{P_k a_{ki}} + b_i .
\end{equation}

The first term in (1) describes the contribution of the investigated process 
that may have a few $\gamma$-lines contributing appreciably to the $i$th bin. 
The parameter $N$ is the number of decays, $\varepsilon_m$ is the detection 
efficiency of the $m$th $\gamma$-line and $a_{mi}$ is the 
contribution of the $m$th line to the $i$th bin. For low-background measurements, a 
$\gamma$-line may be taken to have a gaussian shape. The second term gives 
contributions of background $\gamma$-lines. Here $P_k$ is the area of the 
$k$th $\gamma$-line and $a_{ki}$ is its contribution to the $i$th bin. 
The third term represents the so-called ``continuous background'' 
($b_i$), which has been selected as a straight-line fit after rejecting all peaks 
in the region-of-interest. We select this region as the peak to be 
investigated $\pm$ 30 standard deviations ($\approx$ 20 keV). The likelihood 
function is the product of probabilities for selected bins.  
Normalizing over the parameter $N$ gives the probability density 
function for $N$, which is used to calculate limits for $N$.  To take into 
account errors in the $\gamma$-line shape parameters, peak areas, and other 
factors, one should multiply the likelihood function by the error probability 
distributions for these values and integrate, to provide the average 
probability density function for $N$.

The photon detection efficiency for each investigated process is 
computed using MC simulations.
Finally the lower 
half-life limits are found in the range $(0.4-1.1) \cdot 10^{22}$ yr for the 
transitions (Table~\ref{tab:table1}). 
Table~\ref{tab:table1} also presents other limits on 
these transitions.

\begin{table*}
\caption{\label{tab:table1}Experimental results for $(0\nu+2\nu)\beta\beta$ 
decay of $^{100}$Mo 
to the excited states of $^{100}$Ru. Energy is presented in keV. 
All limits are given at the 90\% C.L.
$^{a)}$Only 0$\nu$ decay mode. 
$^{b)}$Half-life value for $ 2\nu$ decay (see text for the details).}

\begin{tabular}{cccl}
\hline
Excited state & Energy of $\gamma$-rays & 
\multicolumn{2}{c}{$(T^{0\nu+2\nu}_{1/2})_{exp}$ ($10^{20}$ yr)} \\ 
\cline{3-4}
& (efficiency) & this work & other \\ 
&  & & results\\
\hline
$2^+_1 (539.5)$   & 539.5  (4.02\%) & $ > 25 $ & $ > 16 $ \cite{BAR95}  \\
                 &                 &           & 
$ > 1600^{a)} $ \cite{ARN07}  \\
$0^+_1 (1130.3)$  & 539.5  (3.29\%)  & 
$ 7.5 \pm {0.6}(stat) \pm {0.6}(syst)^{b)}$ & 
$  $ see Table 2  \\
                 & 590.8  (3.22\%) &  & $ > 890^{a)}$ \cite{ARN07}  \\  

$2^+_2 (1362.2)$  & 822.6 (1.72\%) & $ > 108$ & $ > 44 $ \cite{KID09}  \\
& 1362.2 (1.34\%)   &   & \\
$0^+_2 (1741.0)$  & 378.9 (1.39\%) & $ > 40$ & $ > 48 $ \cite{KID09} \\
& 1201.5 (1.53\%)  &    & \\
$2^+_3 (1865.1)$  & 734.8  (0.65\%) & $ > 49$ & $ > 43 $ \cite{KID09}  \\
&  1865.1  (0.85\%) &    & \\
$0^+_3 (2051.7)$  & 1512.1  (2.09\%) & $ > 43$ & $ > 40 $ \cite{KID09}  \\
\hline
\end{tabular}
\end{table*}

\section{Discussion} 

Because the technique used in the present work does not allow for a distinction 
between $0\nu\beta\beta$ and $2\nu\beta\beta$ decay, our 
result for double beta decay of $^{100}$Mo to the excited 0$^+_1$ state in 
$^{100}$Ru is the sum of the $0\nu\beta\beta$ and $2\nu\beta\beta$ processes. 
However the detection of only the $2\nu\beta\beta$ decay 
is supported by the following argument - in the recent NEMO-3 paper 
\cite{ARN07} the limit on $0\nu\beta\beta$ decay of $^{100}$Mo to the excited 
0$^+_1$ state was established as $8.9\cdot 10^{22}$ yr, which is two orders of magnitude stronger 
than the half-life value obtained here. 
Therefore, it is safe to assume that our 
result for $T_{1/2}$ refers solely to the $2\nu\beta\beta$ decay.
For the transition to the $0^+_1$ excited state of $^{100}$Ru the obtained value 
is in a good agreement
with results of previous experiments (see Table 2). Our result yields the best statistical accuracy, and one of the smallest systematic errors, of all previous measurements.  As a result the most precise half-life value for 2$\nu$ 
transition of $^{100}$Mo to the 0$^+_1$ excited state of $^{100}$Ru is obtained.

\begin{table}
\caption{Present ``positive" results on $2\nu\beta\beta$ decay of $^{100}$Mo to the first $0^+$
excited state of $^{100}$Ru. $N$ is the number of useful events, $S/B$ is the signal-to-background ratio. 
$^{a)}$ Sum of two peaks. $^{b)}$ Sample was located between two HPGe detectors working in coincidence.
$^{c)}$ The result was obtained using sum spectrum from 4 HPGe detectors. Using coincidence regime half-life 
was measured too, but with a few times worse accuracy.}
\begin{tabular}{lrrrr}
\hline

$T_{1/2}$, yr & $N$ & $S/B$ & Year, Ref. & Method   \\
\hline
$ 6.1^{+1.8}_{-1.1}(stat)\times10^{20}$  & 133$^{a)}$ & $\sim 1/7$ & 1995 \cite{BAR95} & HPGe\\
$9.3^{+2.8}_{-1.7}(stat) \pm 1.4(syst)\times 10^{20}$  & 153$^{a)}$ & $\sim 1/4$ & 1999 \cite{BAR99}& HPGe \\
$6.0^{+1.9}_{-1.1}(stat) \pm 0.6(syst)\times 10^{20}$ & 19.5 & 8/1 & 2001 \cite{DEB01,HOR06} & 2xHPGe$^{b)}$ \\
$5.7^{+1.3}_{-0.9}(stat) \pm 0.8(syst)\times 10^{20}$ & 37.5 & 3/1 & 2007 \cite{ARN07} & NEMO-3 \\
$5.5^{+1.2}_{-0.8}(stat) \pm 0.3(syst)\times 10^{20}$ & 35.5 & 8/1 & 2009 \cite{KID09} & 2xHPGe$^{b)}$ \\
$6.9^{+1.0}_{-0.8}(stat) \pm 0.7(syst)\times 10^{20}$ & 597$^{a)}$ & 1/10 & 2010 \cite{BEL10} & 4xHPGe$^{c)}$ \\
\hline
$7.5 \pm {0.6}(stat) \pm {0.6}(syst)\times 10^{20}$ & 239$^{a)}$ & 2/1 & 2013, & HPGe \\ 
& & & this work \\
\hline
\end{tabular}

\end{table} 

Using the phase space factor value $G = 6.055\cdot 10^{-20}$ yr$^{-1}$ 
\cite{KOT12},  $g_A$ = 1.2701 \cite{BER12} and the measured half-life 
$T_{1/2}=[7.5 \pm {0.6}(stat) \pm {0.6}(syst)]\cdot 10^{20}$ yr, 
we 
obtain a NME value for the $2\nu\beta\beta$ transition to the 
0$^+_1$ excited state of ${\it M_{2\nu}}$(0$^+_1$) = $0.092\pm 0.006$ (scaled by 
the electron rest mass).\footnote{We use here the following relation, 
$T_{1/2}^{-1} = g_A^4\cdot G \cdot M^2_{2\nu}.$ $T_{1/2}$ is the half-life value [yr], 
$G$ is the phase space factor [yr$^{-1}$], $g_A$ is the axial vector coupling constant and $M_{2\nu}$
is the dimensionless nuclear matrix element.}
We can compare this value with the NME value for the $2\nu\beta\beta$ 
transition 
to the ground 
state of $^{100}$Ru, ${\it M}_{2\nu}$(0$^+_{g.s.}$) = 0.1273$^{+0.0038}_{-0.0034}$ (here 
we used the 
average half-life value $T_{1/2}=(7.1\pm 0.4)\cdot 10^{18}$ yr from 
\cite{BAR10}, 
$G$ = $3.308\cdot 10^{-18}$ yr$^{-1}$ from \cite{KOT12} and $g_A$ = 1.2701 from \cite{BER12}). 
Using $G$ values obtained in the framework of the SSD (Single State Dominance)
mechanism \cite{KOT12} we can obtain NME values 
for the transition to the 0$^+_1$ excited state and ground state of $0.089\pm 0.006$ and 
0.1139$^{+0.0034}_{-0.0031}$ respectively. 
Independent of the NME model chosen,
${\it M}_{2\nu}$(0$^+_{g.s.})$ is $\sim$ 25\% greater than ${\it M}_{2\nu}$(0$^+_1$)
with a significance of more than 4$\sigma$. 
The knowledge of these NMEs can be exploited for the deeper understanding of the underlying nuclear structure, 
i.e. for the development of more reliable nuclear structure models. We note that ratio of the half-lives
and the corresponding ratio of NMEs is independent of the value of the axial-vector coupling constant $g_A$. 

The observation of double beta decay to the 2$^+_1$ excited state is rather difficult above the background of 
the decay going to the 0$^+_1$ level. The experimental effect from the 2$^+_1$ decay
is an additional contribution to the peak at an energy of 539.5 keV. In our case 
there is no significant excess. 
Therefore only a limit can be set, 
$T_{1/2} > 2.5\cdot 10^{21}$ yr. This limit can be compared with previous results, 
$> 1.6 \cdot 10^{21}$ yr \cite{BAR95}  and $ > 1.1\cdot 10^{21}$ yr \cite{ARN07}.  
From the general point of view, the decay to the 2$^+$ excited state should be suppressed \cite{HAX84,DOI85}. 
Theoretical values for the half-life of this two neutrino transition usually lay in the 
interval $\sim 10^{23}-10^{26}$ yr \cite{HIR95,STO96,TOI97,RAD07}. Nevertheless 
in Ref. \cite{SUH02} a more ``optimistic" value ($2.1\cdot 10^{21}$ yr) was obtained 
which is however lower than the obtained experimental limit ($2.5\cdot 10^{21}$ yr). 
In the framework of the SSD mechanism, the prediction for this transition 
is $ \sim (1-3)\cdot 10^{23}$ yr \cite{DOM05,SJU08} which is quite 
far from the achieved sensitivity. In the scheme with a ``bosonic" neutrino, the
decay rate can be increased by $\sim$ 100 times \cite{BAR07}. It is therefore interesting 
and important to increase the sensitivity of such measurements.

For the double beta decay of $^{100}$Mo to the 2$^+_1$, 2$^+_2$, 2$^+_3$ and 0$^+_3$ excited 
states of $^{100}$Ru the obtained limits are better than 
the best previous results \cite{KID09} (see Table 1).

\section{Conclusion}

Double beta decay of $^{100}$Mo to the excited states of 
daughter 
nuclei was investigated with a high level of sensitivity. The half-life for 
the 
$2\nu\beta\beta$ decay of $^{100}$Mo to the excited 0$^+_1$ state 
in $^{100}$Ru is measured to be 
$T_{1/2}= [7.5 \pm{0.6}(stat) \pm {0.6}(syst)]\cdot 10^{20}$ yr. 
This is the most precise value yet obtained for this transition.
For other $(0\nu + 2\nu)$ transitions to the
2$^+_1$, 2$^+_2$, 0$^+_2$, 2$^+_3$ and 0$^+_3$ levels in $^{100}$Ru, the obtained limits 
are in the range of 
$(0.25-1.1)\cdot 10^{22}$ yr. The limits for transitions to the 2$^+_1$, 2$^+_2$, 2$^+_3$ and 
0$^+_3$ excited 
states are stronger than the previous results.

\section*{Acknowledgements}
The authors would like to thank the Modane Underground Laboratory staff for 
their technical 
assistance in running the experiment. Portions of this work were supported by 
grants from RFBR (no 12-02-12112 and 13-02-93107) and by grants LG11030 and LM2011027 (MEYS, Czech Republic).





\bibliographystyle{elsarticle-num}
\bibliography{<your-bib-database>}



\end{document}